\documentclass{iaus}
\usepackage{graphicx}

\title[Binary Stars and Globular Cluster Dynamics] 
{Binary Stars and Globular Cluster Dynamics}

\author[Fregeau]   
{John M. Fregeau$^1$}

\affiliation{Northwestern University, Department of Physics and Astronomy, Evanston, IL 60208, USA.
\break email: fregeau@northwestern.edu
\break $^1$Chandra Fellow}

\pubyear{2007}
\volume{246}  
\pagerange{373--}
\date{?? and in revised form ??}
\jname{Proceedings Title IAU Symposium}
\editors{A.C. Editor, B.D. Editor \& C.E. Editor, eds.}
\begin{document}

\maketitle

\begin{abstract}
In this brief proceedings article I summarize the review talk I gave at
the IAU 246 meeting in Capri, Italy, glossing over the well-known results from the literature,
but paying particular attention to new, previously unpublished material.  This new material
includes a careful comparison of the apparently contradictory results of two 
independent methods used to simulate the evolution
of binary populations in dense stellar systems (the direct $N$-body method of 
\cite{2007ApJ...665..707H} and the approximate Monte Carlo method of \cite{2005MNRAS.358..572I}), 
that shows that the two methods may not actually yield contradictory results,
and suggests future work to more directly compare the two methods.
\end{abstract}

\firstsection 
\section{Introduction}

Globular clusters are observed to contain significant numbers of binary star systems---so
many, in fact, that they must have born with binaries (\cite{1992PASP..104..981H}).
Their presence in clusters is important for two complementary reasons.  Through
super-elastic dynamical scattering interactions, they act as an energy source which
may postpone core collapse, and may be the dominant factor in setting the core radii
of observed Galactic globulars.  Similarly, the dense stellar environment and increased
dynamical interaction rate in cluster cores is responsible for the high specific
frequency of stellar ``exotica'' found in clusters, including low-mass X-ray binaries
(LMXBs), cataclysmic variables (CVs), blue straggler stars (BSSs), and recycled millisecond 
pulsars (MSPs).

\section{Evolution of Clusters}\label{sec:evclus}

\subsection{The Negative Heat Capacity of Self-Gravitating Systems}

Imagine finding yourself piloting a spaceship in orbit about a planet.  Your ship is
equipped with rocket thrusters that can either fire in your direction of motion or opposite
it.  If you want to slow down, which way do you fire your thrusters?

The answer is that you fire your thrusters behind you---opposite your direction of motion.
This causes work to be done on your spaceship, which increases your energy, expands your orbit
about the planet, and slows you down.  It's counterintuitive at first, since it's like depressing
the accelerator pedal to slow down, but this behavior is typical of self-gravitating systems, and
is a manifestation of their negative heat capacity---if you add energy to a system it cools down,
if you take energy away from a system is heats up.  A real-life example of this is air drag
on an orbiting satellite, which will actually cause it to speed up.

A basic appreciation of the negative heat capacity of self-gravitating systems goes
a long way in helping to understand the physics of the binary burning phase in clusters
(the phase analogous to the main sequence in stars, in which clusters ``burn'' binaries
instead of hydrogen to support their cores against collapse).  Imagine
a binary star system in a cluster encountering a single star, where the relative
speed between the binary and the single star is smaller than the orbital speed in the
binary system.  When the three stars get close enough to interact strongly, the quickly
moving binary members will tend to transfer some energy to the more slowly moving
incoming single star (energy transfer from hot to cold).  
The result is that when the interaction is over one of the three
stars (and it doesn't have to be the original single star) will leave with a higher
relative velocity than the incoming single star initially had.  Since the binary
system gave up some energy to the single in the interaction it will become more
tightly bound and thus have a larger orbital speed (energy was taken from it
and it got hotter).  The binary we have constructed in this thought
experiment is a ``hard'' binary (since the orbital speed in the binary is larger
than the encounter speed), which clearly becomes harder as a result of dynamical
interactions (\cite{1975MNRAS.173..729H}).  In general, a population of hard primordial
binaries will act as an energy source that supports a cluster's core against collapse through
dynamical scattering interactions (please see \cite{2003gmbp.book.....H} for a more detailed discussion).

It's easy to make an order of magnitude estimate of the importance of binaries in a cluster
(the following discussion closely follows that in \cite{2003gmbp.book.....H}).
Imagine a cluster with $N$ objects, $f_b N$ of which are hard binaries.  Denote the total cluster
mechanical energy as $E_{\rm ext}$, and the total binary binding energy as $E_{\rm int}$.  The
binding energy of a binary with hardness $x$ is then
\begin{equation}
  E_b \equiv xkT \approx x E_{\rm ext} / N \, ,
\end{equation}
where $kT$ represents thermal energy of motion.  The total internal energy is then
\begin{equation}
  E_{\rm int} \approx f_b x E_{\rm ext} \, .
\end{equation}
Since a binary releases energy of order $E_b$ through interactions, binaries are important
when $f_b x \gtrsim 1$.  For example, for binaries of hardness $x=10$ (a reasonable value),
a binary fraction of merely 10\% can be enough to unbind a cluster completely.  It would also
appear that just one sufficiently hard binary could be dynamically very important.  However, it
should be noted that one key element has been left out of the discussion: interaction timescales.
A very hard binary composed of stars that are roughly the average stellar mass in the cluster
would have such a small semimajor axis as to make its interaction time so long that it is 
essentially dynamically irrelevant.

A more detailed analysis of energy generation due to binary burning can give a rough estimate of the
equilibrium core radius in the binary burning phase, and can be compared with observations 
(the following discussion closely follows that in \cite{1989Natur.339...40G}).
In equilibrium the energy generated in the core via binary burning
should equal the energy transported across the half-mass radius
via two-body relaxation.  The binary burning energy generation rate is
\begin{equation}
  \dot E_{\rm bin} \approx n_c (n_c \sigma_{\rm bin} v_c) \left(\frac{4\pi r_c^3}{3}\right) 
  \left(\frac{Gm^2}{2a}\right)
  \sim r_c^3 n_c^2 \frac{G^2 m^3}{v_c} g(f_b, A_{\rm bb}, A_{\rm bs}) \, ,
\end{equation}
where $n_c$ is the core number density, the first term in parentheses is the $n$--$\sigma$--$v$
estimate for the interaction rate of a binary, the second term in parentheses is the core
volume, and the third term is the binding energy of a typical binary.  The function 
$g(f_b, A_{\rm bb}, A_{\rm bs})$ is a dimensionless function of the binary fraction, 
the relative strengths of binary--binary and binary--single energy generation, and is 
of order unity.  The two-body relaxation energy transport rate is
\begin{equation}
  \dot E_{\rm rel} = \frac{|E|}{\alpha t_{\rm rh}} \approx \frac{1}{5 \alpha} \frac{G M^2}{t_{\rm rh} r_h} \, ,
\end{equation}
where $\alpha$ is a constant, $t_{\rm rh}$ is the relaxation time at the half-mass radius,
and $r_h$ is the half-mass radius.  Equating the two expressions yields
\begin{equation}
    \frac{r_c}{r_h} \approx \frac{0.05}{\log_{10}(\gamma N)} g(f_b, A_{\rm bb}, A_{\rm bs}) \, ,
\end{equation}
where the standard expression has been substituted in for the relaxation time, with $\log_{10}(\gamma N)$
the Coulomb logarithm.  For $N=10^6$ this expression yields $r_c/r_h \sim 0.02$ which is in rough agreement
only with the $\sim 20\%$ of Galactic globulars that are observationally classified as core collapsed.

\subsection{Globular Cluster Core Radii}

Recently two independent and very different numerical methods for simulating the evolution of star
clusters have been used to study the core radii of clusters in the binary
burning phase.  One is the direct $N$-body method, which utilizes very few approximations and thus
treats the evolution of clusters on a dynamical (orbital) timescale.
The other is the Monte Carlo method, which uses a number of assumptions in order to treat
the evolution on a relaxation timescale.  To accurately treat them, binary interactions
are handled via direct few-body integration.  Remarkably, the two methods agree quite 
well in the value of $r_c/r_h$ predicted during the binary burning phase 
(\cite{2006MNRAS.368..677H}; \cite{2007ApJ...658.1047F}).  Unfortunately, the value 
predicted by the simulations is at least an order of magnitude smaller than what's observed
for the $\sim 80\%$ of clusters that are observed to be non-core collapsed.  Since the longest
phase of evolution for a cluster is the binary burning phase, it is expected that most
clusters currently observed should be in this phase.  The current state of the field thus
represents a major discrepancy between theory and observations.

Several resolutions to the problem have been proposed.  \cite{2007MNRAS.379...93H},
among others, has noted that there are in fact three different definitions of the
core radius in popular use, with the observational definition possibly being
larger than the standard dynamical definition used in some numerical
codes by a factor of $\sim 4$.  Another suggestion is that there are central
intermediate-mass black holes (IMBHs) in most Galactic globular clusters, which act
as an energy source that would increase core radii to roughly the value
observed (\cite{2006astro.ph.12040T}; \cite{2007MNRAS.tmp..783M}; \cite{2007MNRAS.374..857T}).
It could also be that the ``true'' initial conditions for clusters are of much
higher or lower stellar density than what has traditionally been assumed in simulations
(\cite{2007ApJ...658.1047F}).  Or it could be that stellar mass loss from
enhanced stellar evolution of physical collision products could power
the cores sufficiently (Chatterjee, et al., this volume).
For young clusters, mass segregation of compact
remnants (\cite{2004ApJ...608L..25M}), or the evaporation of the sub-population
of stellar-mass black holes (\cite{2007MNRAS.379L..40M}) could possibly explain the discrepancy.

\subsection{Globular Cluster Binary Fractions}

Observations show that globular clusters currently have core binary
fractions ranging from a few \% (NGC 6397, 47 Tuc, M4), to $\sim 30\%$
(Pal 13, E3, NGC 6752, NGC 288).  The observational techniques used to
determine the core binary fractions are varied, but include observation
of a secondary main sequence, radial velocity studies, and searches for
eclipsing binaries.  All methods involve extrapolation of somewhat
uncertain functions (e.g., the binary mass ratio distribution in 
the first method), although the first method is considered to be the
most complete.  When combined with observations of other cluster
properties, measured core binary fractions enable detailed testing
of cluster evolution models.

Two key processes govern the evolution of the binary fraction in clusters:
binary stellar evolution, and stellar dynamical interactions.  Two codes
currently combine both processes to varying degrees of realism.  The work
of \cite{2007ApJ...665..707H} uses full $N$-body calculations with binary
stellar evolution.  That of \cite{2005MNRAS.358..572I} uses binary 
stellar evolution with a simplified dynamical model that assumes a two-zone cluster
(core and halo) with constant core density, but performs direct few-body integration
of binary scattering interactions.  In perhaps oversimplified terms, \cite{2005MNRAS.358..572I}
find generally that the core binary fraction in clusters decreases with time, requiring
clusters to have been born with large binary fractions to 
explain currently observed binary fractions.
In similarly oversimplified terms, \cite{2007ApJ...665..707H} generally find
that the core binary fraction tends to increase with time.
This is certainly a simplified comparison of the two apparently contradictory results, 
since there are several differences in initial conditions and assumptions used
by the two different methods that need to be taken into account.  The dynamics
and evolution of binary populations in clusters is a complex topic, however,
and the important differences between the two methods and the clusters they model
sometimes get lost in the discussion.

\begin{figure}
\begin{center}
 \includegraphics[width=0.7\textwidth]{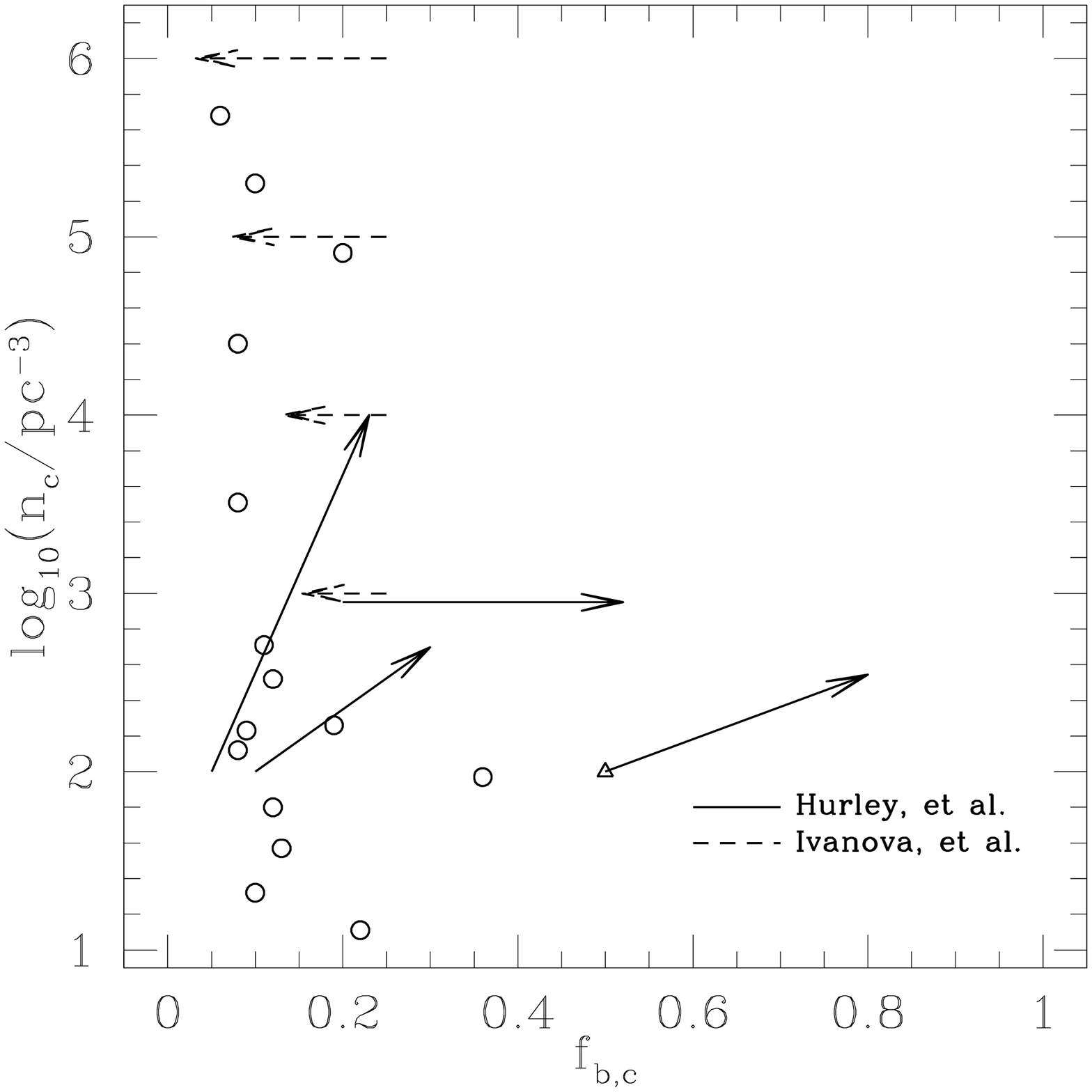}
  \caption{The evolution of cluster evolution models in core number density
    and core binary fraction for the $N$-body simulations of \cite{2007ApJ...665..707H}
    (solid arrows) and the Monte Carlo simulations of \cite{2005MNRAS.358..572I} 
    (dashed arrows).  Note that the binary fraction plotted here is the {\em hard}
    binary fraction.  The models of \cite{2007ApJ...665..707H} use exclusively
    hard binaries, while those of \cite{2005MNRAS.358..572I} start with a wide
    distribution of binaries that extends well into the soft regime.
    The globular clusters for which the core number density
    and binary fraction are known are plotted as open circles.  The single triangle
    point represents a typical open cluster.\label{fig:bincomp}}
\end{center}
\end{figure}

One of the most important parameters governing the evolution of binary fractions
in clusters is the stellar density in the core.  A larger density leads to a higher
dynamical interaction rate, and it is these interactions that can affect stellar evolutionary
processes by modifying orbital parameters of binaries, destroying binaries, exchanging
members of binaries, and creating binaries via tidal capture.  Figure \ref{fig:bincomp}
shows the evolution of cluster evolution models in core number density
and core binary fraction for the $N$-body simulations of \cite{2007ApJ...665..707H}
(solid arrows) and the Monte Carlo simulations of \cite{2005MNRAS.358..572I} 
(dashed arrows).  Note that the binary fraction plotted here is the {\em hard}
binary fraction.  The models of \cite{2007ApJ...665..707H} use exclusively
hard binaries, while those of \cite{2005MNRAS.358..572I} start with a wide
distribution of binaries that extends well into the soft regime.
The globular clusters for which the core number density
and binary fraction are known are plotted as open circles.  The single triangle
point represents a typical open cluster.  In comparing the two models, a clear 
sign of fundamental disagreement would be if an $N$-body arrow and
a Monte Carlo arrow begin in the same region of phase space but point in different
directions.  Only the two arrows at $n_c=10^3\,{\rm pc}^{-3}$ show this disagreement,
pointing in opposite directions.  However, the Monte Carlo method
is less accurate at this low stellar density, since the assumption of constant
core density breaks down here.  Similarly, the $N$-body arrow appears to be horizontal
since the initial and final densities are not known---only the average density is known
for this model.  In other words, the only two contradictory arrows in this diagram may
not faithfully represent their respective methods.  All other arrows are in very different
areas of parameter space, and so unfortunately do not offer direct comparisons between
the two methods.  $N$-body simulations are generally
relegated to clusters with rather low initial densities ($\lesssim 10^3\,{\rm pc}^{-3}$), 
while the Monte Carlo simulations are most accurate
for rather higher densities ($\gtrsim 10^3\,{\rm pc}^{-3}$).  Thus it could very well
be that the results of both methods represent the same underlying evolution.  In order
to fully compare the methods either many more Monte Carlo simulations should be performed
with lower initial densities (although the 
assumption of constant core density breaks down as the density becomes lower), or
$N$-body simulations should be performed for larger initial density and binary
fraction (although this is currently quite computationally expensive).

A viable alternative to properly compare the two methods is to perform an $N$-body
simulation at large initial core density for just a very short time to get
a sense of the direction of evolution in phase space.  It should be noted that the arrows
in Figure \ref{fig:bincomp} are straight-line approximations to the true evolution.  Thus
it should first be tested whether the overall evolution of an $N$-body model
in this parameter space has any relation to the initial, differential evolution.

\section{Evolution of the Binary Population}\label{sec:evbin}

Due to strong binary scattering interactions, globular clusters are home to 
large numbers of ``exotic'' stellar objects, including LMXBs,
CVs, MSPs, and BSSs.  The interactions can create and destroy classes of 
binaries directly through exchange and ionization, and indirectly
by modifying orbital properties and thus affecting binary stellar evolution
processes.

\subsection{The Interaction Frequency for X-Ray Sources}

It was realized over 30 years ago that globular clusters
are overabundant per unit mass in X-ray binaries by orders
of magnitude relative to the disk population 
(\cite{1975ApJ...199L.143C}; \cite{1975Natur.253..698K}).
If X-ray binaries are formed in clusters mainly via binary
scattering interactions, there should be a correlation between
the interaction rate and the number of such binaries in each
cluster (\cite{1987IAUS..125..187V}).  Since the interaction
rate is so heavily used in the literature, it is worthwhile to review its 
derivation.

Imagine a large volume of uniform density $n_1$ of an object
labeled type 1, and similarly for type 2.  The interaction rate
for one member of species 2 with species 1 is
\begin{equation}
  \frac{dN_2}{dt} = n_1 \sigma_{12} v_{12} \, ,
\end{equation}
where $\sigma_{12}$ is the cross section for interaction of an object
of species 1 with an object of species 2, and $v_{12}$ is the relative speed 
between the two species.  The total interaction rate per unit volume
is then
\begin{equation}
  \frac{dN_{\rm int}}{dt\,dV} = n_1 n_2 \sigma_{12} v_{12} \, \, ,
\end{equation}
which is nicely symmetric under transformation between
index 1 and 2.
The total interaction rate for a cluster can be approximated
by multiplying by the core volume, to give
\begin{equation}\label{eq:gamma}
  \Gamma \equiv \frac{dN_{\rm int}}{dt} \approx n_1 n_2 \sigma_{12} v_{12} 
  \frac{4 \pi r_c^3}{3} \propto \rho^2 r_c^3 / v_\sigma \, ,
\end{equation}
where $r_c$ is the core radius, $\rho$ is the core mass density, and $v_\sigma$ 
is the core velocity dispersion.  The last proportionality involves, among other
things, substituting in the mass density and using the gravitational 
focusing dominated interaction cross section.  $\Gamma$, in various 
incarnations, has been used for many years in analyses comparing with
the observed numbers of X-ray sources in clusters.  Recently,
\cite{2003ApJ...591L.131P} have shown that the number of observed
X-ray sources above $4\times 10^{30}\,{\rm erg}/{\rm s}$ in the 0.5--6 keV
range displays a stronger correlation with cluster $\Gamma$ than with any other
cluster parameter.

The fact that the number of X-ray sources in clusters so strongly correlates
with $\Gamma$ is surprising given the number of approximations that go into
deriving it.  For example, the two interacting populations in eq.~(\ref{eq:gamma})
are compact objects (neutron stars and white dwarfs) and stellar binaries,
whose densities may differ dramatically.  The use of $\rho$ in place of 
$n_i$ assumes a proportionality between the stellar density and the
compact object density that is constant among {\em all} clusters.
Furthermore, there is a factor of $f_b$, the binary fraction, that has been
dropped, thereby implicitly assuming that it is constant among all clusters.
There is also a factor of the binary semimajor axis that has been
dropped from the derivation, which may vary since the hard--soft binary
boundary will vary among clusters.
In addition, the recent dynamical history of the cluster may play
an important role.

Improvements to the standard $\Gamma$ analysis have recently been made.
One of the difficulties present in the earlier analyses is that $\Gamma$
strongly correlates with total cluster mass.  This is unfortunate, since 
it makes it more difficult to distinguish dynamically-formed sources
(whose number should correlate with $\Gamma$), from primordial sources
(whose number should correlate with total cluster mass).  An analysis
of the CV populations in clusters using a normalized interaction rate,
$\gamma \equiv \Gamma/M_{\rm clus}$, has shown that like LMXBs, 
CVs are predominantly formed via dynamical encounters (\cite{2006ApJ...646L.143P}).

\section{Summary}\label{sec:summary}

In this proceedings article I have very briefly discussed the connection
between binary stars and globular cluster dynamics, moving from basic
physics to current research in the span of a few paragraphs.  A thorough,
easily readable, and fairly recent discussion of the material can be
found in \cite{2003gmbp.book.....H}.

The primary new material presented here is a comparison in phase space
of the seemingly contradictory binary population evolution simulations 
of \cite{2005MNRAS.358..572I} and \cite{2007ApJ...665..707H}, showing
that they may in fact both represent the same underlying physics.
In other words, new simulations must be performed to better compare the
two very different methods.

\begin{acknowledgments}
For data, stimulating discussions, and general camaraderie, the author thanks
Craig Heinke, Jarrod Hurley, Natasha Ivanova, Frederic Rasio, M. Atakan G\"urkan, 
and Marc Freitag.  JMF acknowledges support from Chandra theory grant TM6-7007X, as well 
as Chandra Postdoctoral Fellowship Award PF7-80047.
\end{acknowledgments}

\begin{discussion}

\discuss{Fregeau}{[at start of presentation] As this conference is an occasion to celebrate Douglas Heggie's 60th birthday, I have done the following to honor him.  For each slide in my presentation for which Douglas had some impact---by directly working on a topic, by influencing the way people think about a topic, or by influencing my own personal thinking on a topic---I have colored the slide background white...  So please keep your eyes peeled for those special slides. [note that all slides from this presentation had white backgrounds]}

\end{discussion}


\begin{thebibliography}{}

\bibitem[Clark(1975)]{1975ApJ...199L.143C} Clark, G.~W.\ 1975, \textit{ApJL}, 199, 
L143

\bibitem[Fregeau \& Rasio(2007)]{2007ApJ...658.1047F} Fregeau, J.~M., \& 
Rasio, F.~A.\ 2007, \textit{ApJ}, 658, 1047 

\bibitem[Goodman \& Hut(1989)]{1989Natur.339...40G} Goodman, J., \& Hut, 
P.\ 1989, \textit{Nature}, 339, 40 

\bibitem[Heggie(1975)]{1975MNRAS.173..729H} Heggie, D.~C.\ 1975, \textit{MNRAS}, 
173, 729 

\bibitem[Heggie \& Hut(2003)]{2003gmbp.book.....H} Heggie, D., \& Hut, P.\ 
2003, The Gravitational Million-Body Problem: A Multidisciplinary Approach 
to Star Cluster Dynamics, by Douglas Heggie and Piet Hut.~ Cambridge 
University Press, 2003, 372 pp.,  

\bibitem[Heggie et al.(2006)]{2006MNRAS.368..677H} Heggie, D.~C., Trenti, 
M., \& Hut, P.\ 2006, \textit{MNRAS}, 368, 677

\bibitem[Hurley(2007)]{2007MNRAS.379...93H} Hurley, J.~R.\ 2007, \textit{MNRAS}, 
379, 93 

\bibitem[Hurley et al.(2007)]{2007ApJ...665..707H} Hurley, J.~R., Aarseth, 
S.~J., \& Shara, M.~M.\ 2007, \textit{ApJ}, 665, 707

\bibitem[Verbunt \& Hut(1987)]{1987IAUS..125..187V} Verbunt, F., \& Hut, 
P.\ 1987, The Origin and Evolution of Neutron Stars, 125, 187

\bibitem[Hut et al.(1992)]{1992PASP..104..981H} Hut, P., et al.\ 1992, 
\textit{PASP}, 104, 981 

\bibitem[Ivanova et al.(2005)]{2005MNRAS.358..572I} Ivanova, N., 
Belczynski, K., Fregeau, J.~M., \& Rasio, F.~A.\ 2005, \textit{MNRAS}, 358, 572

\bibitem[Katz(1975)]{1975Natur.253..698K} Katz, J.~I.\ 1975, \textit{Nature}, 253, 698 

\bibitem[Mackey et al.(2007)]{2007MNRAS.379L..40M} Mackey, A.~D., 
Wilkinson, M.~I., Davies, M.~B., \& Gilmore, G.~F.\ 2007, \textit{MNRAS}, 379, L40

\bibitem[Merritt et al.(2004)]{2004ApJ...608L..25M} Merritt, D., Piatek, 
S., Portegies Zwart, S., \& Hemsendorf, M.\ 2004, \textit{ApJL}, 608, L25

\bibitem[Miocchi(2007)]{2007MNRAS.tmp..783M} Miocchi, P.\ 2007, \textit{MNRAS}, 783 

\bibitem[Pooley et al.(2003)]{2003ApJ...591L.131P} Pooley, D., et al.\ 
2003, \textit{ApJL}, 591, L131

\bibitem[Pooley \& Hut(2006)]{2006ApJ...646L.143P} Pooley, D., \& Hut, P.\ 
2006, \textit{ApJL}, 646, L143

\bibitem[Trenti(2006)]{2006astro.ph.12040T} Trenti, M.\ 2006, ArXiv 
Astrophysics e-prints, arXiv:astro-ph/0612040

\bibitem[Trenti et al.(2007)]{2007MNRAS.374..857T} Trenti, M., Ardi, E., 
Mineshige, S., \& Hut, P.\ 2007, \textit{MNRAS}, 374, 857

\end{thebibliography}
\end{document}